\documentclass[aps,prl,final,10pt,twocolumn,showpacs,superscriptaddress,footinbib]{revtex4-1}
\usepackage{amsmath,amssymb}
\usepackage{graphicx,color}
\usepackage[squaren]{SIunits}
\usepackage[english]{babel}

\begin{document}
\title{Circular motion of asymmetric self-propelling particles}

\author{Felix K{\"u}mmel}
\affiliation{2.\ Physikalisches Institut, Universit\"at Stuttgart, D-70569 Stuttgart, Germany} 

\author{Borge ten Hagen}
\affiliation{Institut f{\"u}r Theoretische Physik II, Weiche Materie,
Heinrich-Heine-Universit{\"a}t D{\"u}sseldorf, D-40225 D{\"u}sseldorf, Germany}

\author{Raphael Wittkowski}
\affiliation{School of Physics and Astronomy, University of Edinburgh, Edinburgh, EH9 3JZ, United Kingdom}

\author{Ivo Buttinoni}
\affiliation{2.\ Physikalisches Institut, Universit\"at Stuttgart, D-70569 Stuttgart, Germany} 

\author{Ralf~Eichhorn} 
\affiliation{Nordita, Royal Institute of Technology, and Stockholm  
University, SE-10691 Stockholm, Sweden} 

\author{Giovanni Volpe}
\affiliation{2.\ Physikalisches Institut, Universit\"at Stuttgart, D-70569 Stuttgart, Germany} 
\affiliation{Present address: Department of Physics, Bilkent University, Cankaya,
Ankara 06800, Turkey}

\author{Hartmut L{\"o}wen}
\affiliation{Institut f{\"u}r Theoretische Physik II, Weiche Materie,
Heinrich-Heine-Universit{\"a}t D{\"u}sseldorf, D-40225 D{\"u}sseldorf, Germany}

\author{Clemens Bechinger}
\affiliation{2.\ Physikalisches Institut, Universit\"at Stuttgart, D-70569 Stuttgart, Germany} 
\affiliation{Max-Planck-Institut f\"ur Intelligente Systeme, D-70569 Stuttgart, Germany}

\date{\today}

\begin{abstract}
Micron-sized self-propelled (active) particles can be considered as
model systems for characterizing more complex biological organisms
like swimming bacteria or motile cells. We produce asymmetric microswimmers by soft lithography and study their circular motion on a substrate and near channel boundaries. Our experimental observations are in full agreement with a theory of Brownian dynamics for asymmetric self-propelled particles, which couples their translational and orientational motion.
\end{abstract}


\maketitle


Micron-sized particles undergoing active Brownian motion \cite{schimanskyreview} currently receive considerable attention from experimentalists and theoreticians because their locomotion behavior resembles the trajectories of motile microorganisms \cite{EbbensH2010,swimmers2,swimmers3,catesreview}. Therefore, such systems allow interesting insights into how active matter \cite{marchettireview} organizes into complex dynamical structures. During the last decade, different experimental realizations of microswimmers have been investigated, where, e.g., artificial flagella \cite{DreyfusBRFSB2005} or thermophoretic \cite{thermo} and diffusiophoretic \cite{diffusio} driving forces lead to active motion of micron-sized objects. So far, most studies have concentrated on spherical or rod-like microswimmers whose dynamics is well described by a persistent random walk with a transition from a short-time ballistic to a long-time diffusive behavior \cite{golestanian}. 
Such simple rotationally symmetric shapes, however, usually provide only a crude approximation for self-propelling microorganisms, which are often asymmetric around their propulsion axis. Then, generically, a torque is induced that significantly perturbs the swimming path and results in a characteristic circular motion. 

In this Letter, we experimentally and theoretically study the motion of asymmetric self-propelled particles in a viscous liquid. We observe a pronounced circular motion whose curvature radius is independent of the propulsion strength but only depends on the shape of the swimmer. Based on the shape-dependent particle mobility matrix, we propose two coupled Langevin equations for the translational and rotational motion of the particles under an intrinsic force, which dictates the swimming velocity. The anisotropic particle shape then generates an additional velocity-dependent torque, in agreement with our measurements. Furthermore, we also investigate the motion of asymmetric particles in lateral confinement. In agreement with theoretical predictions we find either a stable sliding
along the wall or a reflection, depending on the contact angle. 

\begin{figure}[t]
\centering
\includegraphics[width = \columnwidth]{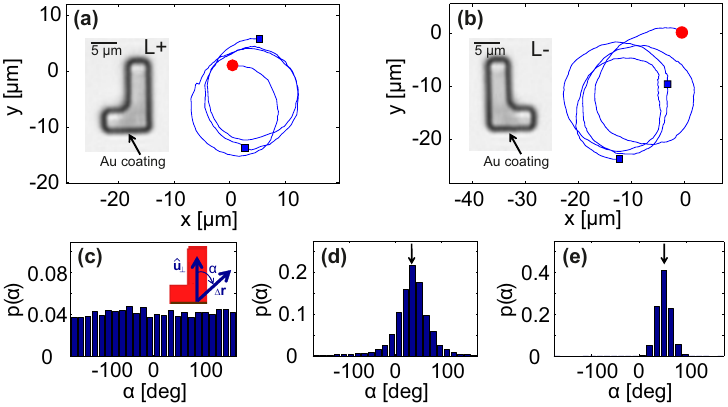}
\caption{\label{Fig.1}(Color online) (a),(b) Trajectories of an (a) L+ and (b) L-- swimmer for an illumination intensity of \unit{7.5}{\micro\watt/\micro\metre^{2}}. (Red) bullets and (blue) square symbols correspond to  initial particle positions and those after \unit{1}{\minute} each, respectively. The insets show microscope images of two different swimmers with the Au coating (not visible in the brightfield image) indicated by an arrow. (c),(d),(e) Probability distributions $p(\alpha)$ of the angle $\alpha$ (see inset in (c)) between the normal vector $\hat{\mathbf{u}}_{\perp}$ of the metal coating and the displacement vector $\Delta \mathbf{r}$ of an L+ particle in time intervals of $\unit{12}{\second}$ each  for illumination intensities (c) I = \unit{0}{\micro\watt/\micro\metre^{2}}, (d) \unit{5}{\micro\watt/\micro\metre^{2}}, and (e) \unit{7.5}{\micro\watt/\micro\metre^{2}}.}
\end{figure}

Asymmetric L-shaped swimmers with arm lengths of $9$ and \unit{6}{\micro\metre} were fabricated from photoresist SU-8 by photolithography \cite{photolit}. In short, a \unit{2.5}{\micro\metre} thick layer of SU-8 is spin coated onto a silicon wafer, soft-baked for \unit{80}{\second} at \unit{95}{\degreecelsius} and then exposed to ultraviolet light through a photo mask. After a post-exposure bake at \unit{95}{\degreecelsius} for \unit{140}{\second} the entire wafer with the attached particles is coated  with a \unit{20}{\nano\meter} thick Au layer by thermal evaporation. When the wafer is tilted to approximately \unit{90}{\degree} relative to the evaporation source, the Au is selectively deposited at the front side of the short arms as schematically shown in Figs.\ \ref{Fig.1}(a),(b). Finally, the coated particles are released from the wafer by an ultrasonic bath treatment. A small amount of L-shaped particles is suspended in a homogeneous mixture of water and 2,6-lutidine at critical 
concentration ($28.6$ mass percent of lutidine), which is kept several degrees below its lower critical point ($T_{\mathrm{C}} = \unit{34.1}{\degreecelsius}$) \cite{lutidine}. To confine the particle's motion to two dimensions, the suspension is contained in a sealed sample cell with \unit{7}{\micro\metre} height. The particles are localized above the lower wall at an average height of about \unit{100}{\nano\meter} due to the presence of electrostatic and gravitational forces. Under these conditions, they cannot rotate between the two configurations shown in Figs.\ \ref{Fig.1}(a),(b), which will be denoted as L+ (left) and L-- (right) in the following. When the sample cell is illuminated by light ($\lambda = \unit{532}{\nano\metre}$) with intensities ranging on the order of several $\unit{}{\micro\watt/\micro\metre^{2}}$, the metal cap becomes slightly heated above the critical point and thus induces a local demixing of the solvent \cite{chemo1,VolpeBVKB2011}. This leads to a self-phoretic particle motion similar to what has been observed in other systems \cite{PalacciCBYB2010,paxton,vicario}.

\begin{figure}[t]
\includegraphics[width=\columnwidth]{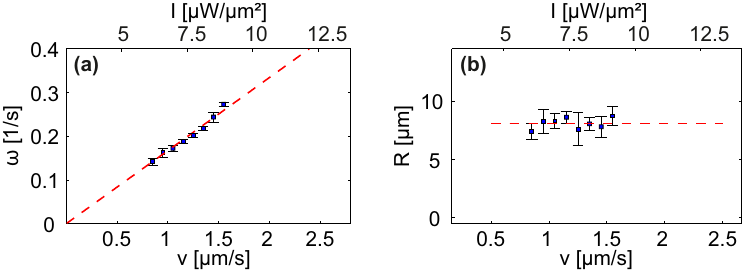}
\caption{\label{Fig.2}(Color online) (a) Angular velocity $\omega$ and (b)
radius $R$ of the circular motion of an L+ swimmer plotted as
functions of the linear velocity $v=\lvert\mathbf{v}\rvert$ and the illumination intensity $I\!\sim\! v$. 
The dashed lines correspond to a linear fit with nonzero and zero slope, respectively.}
\end{figure}

Figures \ref{Fig.1}(a),(b) show trajectories of L+
and L-- swimmers obtained by digital video microscopy for an illumination intensity of
\unit{7.5}{\micro\watt/\micro\metre^{2}}, which corresponds to a mean propulsion speed of \unit{1.25}{\micro\metre/\second}. 
In contrast to spherical swimmers, here a pronounced
circular motion with clockwise (L+) and counter-clockwise (L--)
direction of rotation is observed. For the characterization of trajectories we determined the center-of-mass position $\mathbf{r}(t)=(x(t),y(t))$ and the normalized orientation vector $\hat{\mathbf{u}}_{\perp}$ of the particles (see inset of  Fig.\ \ref{Fig.1}(c)). From this, we derived the angle $\alpha$ between the displacement vector $\Delta \mathbf{r}$ and the particle's body orientation $\hat{\mathbf{u}}_{\perp}$. Figures \ref{Fig.1}(c)-(e) show how the normalized probability distribution $p(\alpha)$ changes with increasing illumination intensity $I$. 
In case of pure Brownian motion (see Fig.\ \ref{Fig.1}(c)) $p(\alpha)\approx \mathrm{const.}$ since the orientational and translational degrees of freedom are decoupled when only random forces are acting on the particle. In presence of a propulsion force which is constant in the body frame of the particle, however, the translational and  rotational motion of an asymmetric particle are coupled. This leads to a peaked behavior of $p(\alpha)$ as shown in Figs.\ \ref{Fig.1}(d),(e). The peak's halfwidth decreases with increasing illumination intensity since the contribution of the Brownian motion is more and more dominated by the propulsive part. In addition, the peaks are shifted to positive (negative) values for a particle swimming in (counter-)clockwise direction. The position of the peak is given by $\alpha=\pi \Delta t / \tau$, where $\tau$ is the intensity-dependent cycle duration of the circle swimmer (cf., Fig.\ \ref{Fig.2}(b)) and $\Delta t$ is the considered time interval. This estimate (see arrows in Figs.\ \ref{Fig.1}(d) ($\tau=\unit{60}{\second}$) and \ref{Fig.1}(e) ($\tau=\unit{40}{\second}$)) is in good agreement with the experimental data. 
The shift of the maximum of $p(\alpha)$ documents a torque responsible for the observed circular motion of such asymmetric swimmers.
In contrast to an externally applied constant torque \cite{giovanni}, here it is due to viscous forces acting on the self-propelling particle. This is supported by the experimental observation that the particle's angular velocity $\omega(t)=\mathrm{d}\alpha/\mathrm{d}t$ increases linearly with its total translational velocity $v(t)$ (see Fig.\ \ref{Fig.2}(a)).
As a direct result of the linear relationship between $\omega$ and $v$, the radius $R$ of the circular trajectories becomes independent of the propulsion speed, which is set by the illumination intensity (see Fig.\ \ref{Fig.2}(b)).

\begin{figure}[t]
\includegraphics[width=\columnwidth]{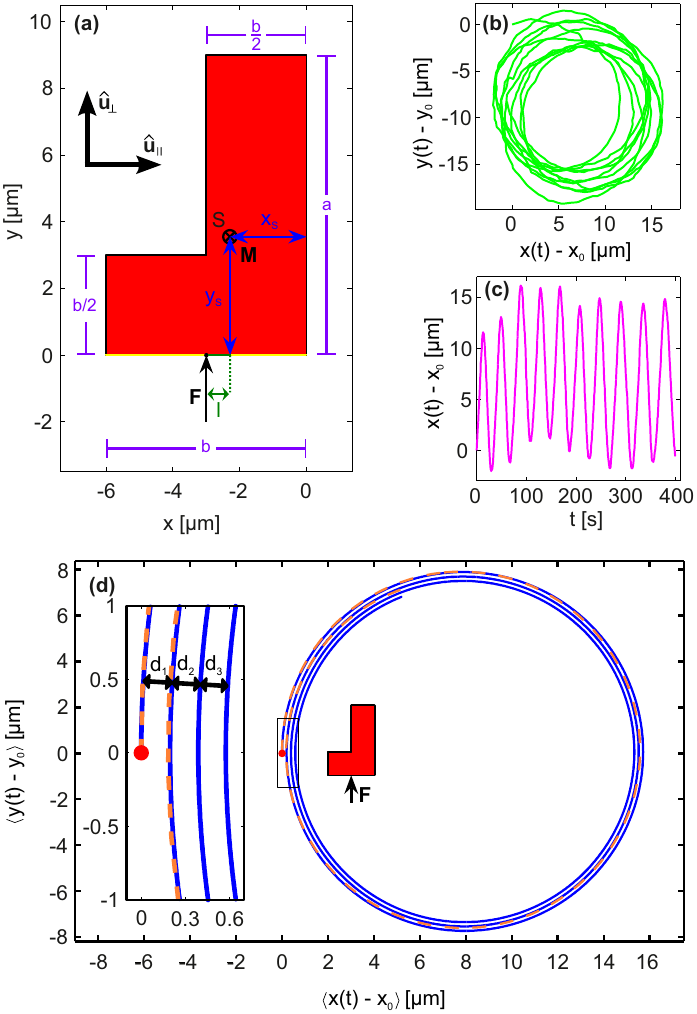}
\caption{\label{Fig.3}(Color online) (a) Geometrical sketch of an ideal L+ swimmer as considered in our model. The dimensions are $a=\unit{9}{\micro\metre}$, $b=\unit{6}{\micro\metre}$, $x_{\mathrm{S}}=\unit{2.29}{\micro\metre}$, and $y_{\mathrm{S}}=\unit{3.55}{\micro\metre}$ (for homogeneous mass density and an additional \unit{20}{\nano\meter} thick Au layer). The internal force $\mathbf{F}$ induces a torque $\mathbf{M}$ on the center-of-mass $\mathrm{S}$ depending on the lever arm $l$. (b),(c) Visualization of the experimental trajectory (for an illumination intensity of $I=\unit{7.5}{\micro\watt/\micro\metre^{2}}$) that is used for the quantitative analysis of the fluctuation-averaged trajectory in (d). The dashed curve in (d) is the experimental one, and the solid curve shows the
theoretical prediction with the starting point indicated by a red bullet. Inset: close-up of the framed area in the plot.}
\end{figure}

For a theoretical description of the motion of asymmetric swimmers, we consider an effective propulsion force $\mathbf{F}$ \cite{tenHagenvTL2011}, which is constant in a
body-fixed coordinate system that rotates with the active particle. 
With the unit  vectors $\hat{\mathbf{u}}_{\perp}=(-\sin\phi,\cos\phi)$ and $\hat{\mathbf{u}}_{\parallel}=(\cos\phi,\sin\phi)$ (see Figs.\ \ref{Fig.1}(c) and \ref{Fig.3}(a)), where -- in case of L-shaped particles -- $\phi$ is the angle between the short arm and the $x$ axis, the propulsion force $\mathbf{F}$ is given by $\mathbf{F}=F \hat{\mathbf{u}}_\mathrm{int}$ with $\hat{\mathbf{u}}_\mathrm{int}=(c \hat{\mathbf{u}}_{\parallel} + \hat{\mathbf{u}}_{\perp})/\sqrt{1+c^2}$ with the constant $c$ depending on how the force is aligned relative to the particle shape. If the propulsion force is aligned along the long axis
$\hat{\mathbf{u}}_{\perp}$, one obtains $c=0$, i.e., $ \hat{\mathbf{u}}_{\mathrm{int}} = \hat{\mathbf{u}}_{\perp}$. In case of an asymmetric particle, the propulsion force leads also to a velocity-dependent torque relative to the particle's center-of-mass. For $c=0$ this torque is given by  $M=l F$ with $l$ the effective lever arm (see Fig.\ \ref{Fig.3}(a)). 
Our theoretical model is valid for arbitrary particle shapes and values of $c$ and $l$. However, for the sake of clarity, we set $c=0$ as this applies for the L-shaped particles considered here. 
Accordingly, we obtain the following coupled Langevin equations, which describe the motion of an asymmetric microswimmer
\begin{equation}%
\begin{split}%
\dot{\mathbf{r}}&= \beta F \big(\mathrm{D}_{\mathrm{T}}\hat{\mathbf{u}}_{\perp}+l\mathbf{D}_{\mathrm{C}} \big)
+\boldsymbol{\zeta}_{\mathbf{r}} \;,\\
\dot{\phi}&= \beta F\big(l D_{\mathrm{R}} +\mathbf{D}_{\mathrm{C}}\!\cdot\!\hat{\mathbf{u}}_{\perp}\big) + \zeta_{\phi} \;.
\end{split}%
\label{eq:LangevinGLG}%
\end{equation}%
Here, $\beta=1/(k_{\mathrm{B}}T)$ is the inverse effective thermal energy of the system. These Langevin equations contain the
translational short-time diffusion tensor
$\mathrm{D}_{\mathrm{T}}(\phi)=\,D_{\parallel}\hat{\mathbf{u}}_{\parallel}\otimes\hat{\mathbf{u}}_{\parallel}
+D^{\perp}_{\parallel}(\hat{\mathbf{u}}_{\parallel}\otimes\hat{\mathbf{u}}_{\perp}+\hat{\mathbf{u}}_{\perp}\!\otimes\hat{\mathbf{u}}_{\parallel})
+D_{\perp}\hat{\mathbf{u}}_{\perp}\!\otimes\hat{\mathbf{u}}_{\perp}$
with the dyadic product $\otimes$ and the translation-rotation
coupling vector
$\mathbf{D}_{\mathrm{C}}(\phi)=D^{\parallel}_{\mathrm{C}}\hat{\mathbf{u}}_{\parallel}+D^{\perp}_{\mathrm{C}}\hat{\mathbf{u}}_{\perp}$ \footnote{Alternatively, it is also possible to use the resistance matrix formalism \cite{Lauga_review:09,Purcell:77}.}.
The translational diffusion coefficients $D_{\parallel}$, $D^{\perp}_{\parallel}$,
and $D_{\perp}$, the coupling coefficients $D^{\parallel}_{\mathrm{C}}$ and
$D^{\perp}_{\mathrm{C}}$, and the rotational diffusion constant $D_{\mathrm{R}}$ are determined by the
specific shape of the particle. Finally, $\boldsymbol{\zeta}_{\mathbf{r}}(t)$
and $\zeta_{\phi}(t)$ are Gaussian noise terms of zero mean and
variances
$\langle\boldsymbol{\zeta}_{\mathbf{r}}(t_{1})\otimes\boldsymbol{\zeta}_{\mathbf{r}}(t_{2})\rangle
=2\:\!\mathrm{D}_{\mathrm{T}}\:\!\delta(t_{1}-t_{2})$,
$\langle\boldsymbol{\zeta}_{\mathbf{r}}(t_{1})\:\!\zeta_{\phi}(t_{2})\rangle
=2\:\!\mathbf{D}_{\mathrm{C}}\:\!\delta(t_{1}-t_{2})$, and
$\langle\zeta_{\phi}(t_{1})\:\!\zeta_{\phi}(t_{2})\rangle=2\:\!D_{\mathrm{R}}\:\!\delta(t_{1}-t_{2})$
\footnote{Due to the multiplicative noise, an additional drift
term has to be taken into account, when Eqs.\
\eqref{eq:LangevinGLG} are solved numerically.}.

In case of vanishing noise, Eq.\ \eqref{eq:LangevinGLG} immediately leads to a circular
trajectory with radius
\begin{equation}
R = \sqrt{\frac{(D^{\perp}_{\parallel} + l
D^{\parallel}_{\mathrm{C}})^2+(D_{\perp} + l D^{\perp}_{\mathrm{C}})^2}{(D^{\perp}_{\mathrm{C}} +
l D_{\mathrm{R}})^2}} \;.
\label{eq:Radiussemi}%
\end{equation}
In agreement with the experimental observation (see Fig.\ \ref{Fig.2}(b)) the radius does not depend on the particle velocity set by the propulsion force. Rather, the value of $R$ is only determined by the particle's geometry, which defines its diffusional properties.
Moreover, the translational and angular particle
velocities are $v = \beta F \sqrt{(D^{\perp}_{\parallel} + l
D^{\parallel}_{\mathrm{C}})^2+(D_{\perp} + l D^{\perp}_{\mathrm{C}})^2}$
and
$\omega = \beta F (D^{\perp}_{\mathrm{C}} +
l D_{\mathrm{R}})$.
Both quantities are proportional to the internal force $F$ and ensure
$R=v/\lvert\omega\rvert$ in perfect agreement with the experimental results shown in Fig.\ \ref{Fig.2}(a). 

For a quantitative comparison with the experimental data, most importantly,
the diffusion and coupling coefficients for the particles under study have to
be determined. They constitute the components of the generalized diffusion
matrix and are, in principle, obtained from solving the Stokes equation that describes
the low Reynolds number flow field around a particle close to the substrate \cite{HappelB1991}.
This procedure can be approximated by using a bead model \cite{Carrasco99},
where the L-shaped particle is assembled from a large number of rigidly
connected small spheres. Exploiting the linearity of the Stokes equation,
the hydrodynamic interactions between any pair of those beads can be
superimposed to calculate the generalized mobility tensor of the L-shaped
particle and from that its diffusion and coupling coefficients; details
of the calculation are outlined in Ref.\ \cite{Carrasco99}. This method is well established
for arbitrarily shaped particles in bulk solution \cite{Carrasco99,delaTorreNMDC1994}. We
take into account the presence of the substrate by using the Stokeslet
close to a no-slip boundary \cite{Blake71} to model the hydrodynamic
interactions between the component beads in the bead model.
For the L-shaped particles considered here, we find that the value of $D_\perp$ exceeds the terms $D^{\perp}_{\parallel}$, $l D^{\parallel}_{\mathrm{C}}$, and $l D^{\perp}_{\mathrm{C}}$ in the numerator of Eq.\
\eqref{eq:Radiussemi} by more than one order of magnitude (given that $l$ is in the range of $\unit{1}{\micro\metre}$). On the other hand, the value of $D^{\perp}_{\mathrm{C}}$ is negligible compared to $lD_R$. This finally yields 
\begin{equation}
R = \lvert D_\perp / (l D_\mathrm{R})\rvert 
\label{eq:appRadius}%
\end{equation}
as an approximate expression for the trajectory radius and, correspondingly, 
\begin{equation}
\omega = \beta D_\mathrm{R} l F
\label{eq:appomega}%
\end{equation}
for the angular velocity. 

\begin{table}[ht]
\caption{\label{tab:D}Diffusion coefficients for the L-shaped particle in Fig.\ \ref{Fig.3}(a) on a substrate: translational diffusion along the long ($D_\perp$) and the short ($D_\parallel$) axis of the L-shaped particle as well as rotational diffusion constant $D_\mathrm{R}$.}
\begin{ruledtabular}
\begin{tabular}{lcc}
 & experiment& theory  \\

\hline
$D_{\perp}\,\,[10^{-3} \micro \squaren \metre\reciprocal\second]$   &         $8.1 \pm 0.6$    & $8.3$          \\
$D_{\parallel}\,\,[10^{-3} \micro \squaren \metre\reciprocal\second]$          &       $7.2 \pm 0.4$         & $7.5$            \\
$D_{\mathrm{R}}\,\,[10^{-4} \reciprocal\second]$                          &           $6.2 \pm 0.8$       &  $6.1$    \\
\end{tabular}
\end{ruledtabular}
\end{table}

We determined the diffusion coefficients $D_\perp$, $D_\parallel$, and $D_\mathrm{R}$ experimentally under equilibrium conditions (i.e., in the absence of propulsion) from the short-time correlations of the translational and orientational components of the particle's trajectories \cite{Han:06,Han:09} (see Tab.\ \ref{tab:D}). The experimental values are in good agreement with the theoretical predictions.

Inserting the experimentally determined values for the diffusion coefficients and the mean trajectory radius $R = \unit{7.91}{\micro\metre}$ into Eq.\ \eqref{eq:appRadius}, we obtain the effective lever arm $l=\unit{-1.65}{\micro\metre}$. This value is about a factor of two larger compared to an ideally shaped L-particle (see Fig.\ \ref{Fig.3}(a)) with its  propulsion force perfectly centered at the middle of the Au layer. This deviation suggests that the force is shifted by $\unit{0.94}{\micro\metre}$ in lateral direction, which is most likely caused by small inhomogeneities of the Au layer due to shadowing effects during the grazing incidence metal evaporation.
Accordingly, from Eq.\ \eqref{eq:appomega} we obtain the intensity-dependent propulsion force $F/I=\unit{0.83 \times 10^{-13}}{\newton \micro\metre^{2}/\micro\watt}$.

To compare the trajectories obtained from the Langevin equations \eqref{eq:LangevinGLG} with experimental data, we divided the measured trajectories into smaller segments and superimposed them such that the initial slopes and positions of the segments overlap. After averaging the data we obtained the noise-averaged mean swimming path, which is predicted to be a logarithmic spiral 
(\emph{spira mirabilis}) \cite{vanTeeffelenL2008} that is given in polar coordinates by
\begin{equation}
r(\phi)=\beta F \sqrt{\frac{D_\perp^2}{D_\mathrm{R}^2+\omega^2}}
\exp\!\bigg(\!-\frac{D_\mathrm{R}}{\omega}(\phi-\phi_{0})\!\bigg) .
\label{spira}
\end{equation} 
Qualitatively, such spirals can be understood as follows: in the absence of thermal noise, the average swimming path corresponds to a circle with radius $R$ given by Eq.\ \eqref{eq:appRadius}. In the presence of thermal noise, however, single trajectory segments become increasingly different as time proceeds. This leads to decreasing distances $d_{i}$  between adjacent turns of the mean swimming path ($d_i/d_{i+1}=\exp{(2\pi D_\mathrm{R} /\lvert \omega \rvert)}$, see Fig.\ \ref{Fig.3}(d)) and, finally, to the convergence in a single point for $t \to \infty$. Due to the alignment of the initial slope, this point is shifted relative to the starting point depending on the alignment angle and the circulation direction of the particle.

The solid curve in Fig.\ \ref{Fig.3}(d) is the theoretical prediction (see Eq.\ \eqref{spira}) with the measured values of $D_\perp$, $D_\mathrm{R}$, and $\omega$. On the other hand, the dashed curve in Fig.\ \ref{Fig.3}(d) visualizes the noise-averaged trajectory determined directly from the experimental data (see Figs.\ \ref{Fig.3}(b),(c)). The agreement of the two curves constitutes a direct verification of our theoretical model on a fundamental level.

\begin{figure}[t]
\includegraphics[width=\columnwidth]{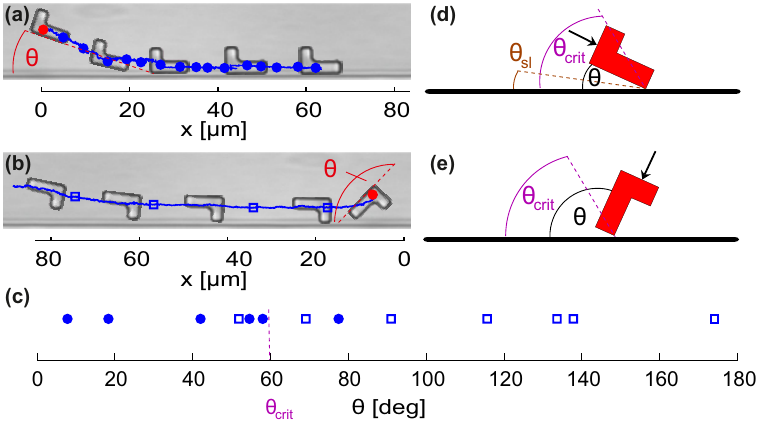}
\caption{\label{Fig.4}(Color online) (a),(b) Trajectories of an 
L+ swimmer approaching a straight wall at different angles (symbols correspond
to positions after \unit{1}{\minute} each). (c) Experimentally determined particle motion for different contact angles $\theta$. Bullets and open squares correspond to particle sliding and reflection. (d),(e) Visualization of 
the predicted types of motion for an L+ 
swimmer with arrows indicating the direction of the propulsion force: (d) stable sliding and (e) reflection. 
The angles are defined in the text.}
\end{figure}

Finally, we also address the motion of asymmetric swimmers under confinement, e.g., their
interaction with a straight wall. This is shown in Fig.\ \ref{Fig.4}(a) exemplarily for an L+ swimmer which approaches the wall at an angle $\theta$. Due to the internal torque associated with the active particle motion, it becomes stabilized at the wall and smoothly glides to the right along the interface. In contrast, for a much larger initial contact angle the internal torque rotates the front part of the particle away
from the obstacle, the motion is unstable, and the swimmer is reflected by the wall (see Fig.\ \ref{Fig.4}(b)) \footnote{Note that this effect occurring for hard colloidal swimmers is different from the 
situation of \emph{Escherichia coli} bacteria confined in the proximity of a surface \cite{LaugaDWS2006}. In the latter case, the circular motion is generated by the rotation of the body and the flagella, coupled with the hydrodynamic interactions with the boundary.}. Figure \ref{Fig.4}(c) shows the observed dependence of the motional behavior as a function of the approaching angle.

The experimental findings are in line with an instability 
analysis based on
a torque balance condition 
of an L-shaped particle at wall contact as a 
function of its
contact  angle $\theta$. 
For $\theta_{\mathrm{crit}} < \theta < \pi$ (see Figs.\ \ref{Fig.4}(b),(e)) with a critical angle $\theta_{\mathrm{crit}}$, the particle is reflected, while for  $0<\theta<\theta_{\mathrm{crit}}$ (see Figs.\ \ref{Fig.4}(a),(d)) stable sliding with an angle 
$\theta_\mathrm{sl}$ occurs. 
Both, $\theta_\mathrm{sl}$ and
$\theta_{\mathrm{crit}}$ are given as stable and unstable solutions, respectively, of 
the torque balance condition
\begin{equation}
\lvert l \rvert = \left[ \left (a-y_\mathrm{S}\right) \cos\theta - 
x_\mathrm{S} \sin\theta\right] \sin\theta \;.
\label{phicrit}
\end{equation}
For $l=\unit{-0.71}{\micro\metre}$ corresponding to an ideal L-shaped particle with the propulsion force centered in the middle of the Au layer, we obtain $\theta_\mathrm{sl}=\unit{8.0}{\degree}$ and $\theta_\mathrm{crit}=\unit{59.2}{\degree}$, which is in good agreement with the measured value of about  $\theta_\mathrm{crit}=\unit{60}{\degree}$ (see Fig.\ \ref{Fig.4}(c)). 
The observed scatter in the experimental data around the critical angle is due to thermal fluctuations that wash out the sharp transition between the sliding and the reflection regime.

In conclusion, we have
demonstrated that due to viscous forces of the surrounding liquid, asymmetric microswimmers are subjected to a velocity-dependent torque.
This leads to a circular motion, which is observed in experiments in agreement with a theoretical model
based on two coupled Langevin equations. In a channel geometry, this torque leads either to a reflection or  a stable sliding motion along the wall. An interesting question  for the future addresses how  asymmetric swimmers move through patterned media. In the presence of a drift force, one may expect Shapiro steps in the particle current similar to what has  also been found in colloidal systems driven  by a circular drive \cite{reichhardt}. Another appealing outlook addresses the motion of chiral swimmers in the presence of external fields such as gravity \cite{WittkowskiL2012}. In case of asymmetric particles, this leads to an orientational alignment during their sedimentation, which may result in a preferential motion relative to gravity similar to the gravitactic behavior of asymmetric cells  as, e.g., \emph{Chlamydomonas} \cite{gravitaxis1,gravitaxis2}.

\begin{acknowledgments}
We thank M. Aristov for assistance in particle preparation and M.\ Heinen for helpful discussions.
This work was supported by the DFG within SPP 1296 and SFB TR6-C3 as well as by the
Marie Curie-Initial Training Network Comploids funded by the European Union Seventh Framework Program (FP7). 
R.\ W.\ gratefully acknowledges financial support from a Postdoctoral Research Fellowship (WI 4170/1-1) of the DFG. 
\end{acknowledgments}

\bibliography{refs}

\end{document}